# Optical data transmission at 44Tb/s and 10 bits/s/Hz over the C-band with standard fibre and a single micro-comb source

Bill Corcoran[1], Mengxi Tan,[2] Xingyuan Xu,[2] Andreas Boes,[3], Jiayang Wu,[2]
Thach G. Nguyen,[3] Sai T. Chu,[4] Brent E. Little,[5] Roberto Morandotti,[6] Arnan Mitchell,[3] and David J. Moss[2]

[1] *Department of Electrical and Computer System Engineering, Monash University, Clayton, VIC 3168 Australia*
[2] *Optical Sciences Centre, Swinburne University of Technology, Hawthorn, VIC 3122, Australia*
[3] *School of Engineering, RMIT University, Melbourne, VIC 3001, Australia*
[4] *Department of Physics and Material Science, City University of Hong Kong, Tat Chee Avenue, Hong Kong, China.*
[5] *Xi'an Institute of Optics and Precision Mechanics Precision Mechanics of CAS, Xi'an, China.*
[6] *INRS-Énergie, Matériaux et Télécommunications, 1650 Boulevard Lionel-Boulet, Varennes, Québec, J3X 1S2, Canada.*

**Micro-combs [1-4] - optical frequency combs generated by integrated micro-cavity resonators – offer the full potential of their bulk counterparts [5,6], but in an integrated footprint. The discovery of temporal soliton states (DKS – dissipative Kerr solitons) [4,7-11] as a means of mode-locking micro-combs has enabled breakthroughs in many fields including spectroscopy [12,13], microwave photonics [14], frequency synthesis [15], optical ranging [16,17], quantum sources [18,19], metrology [20,21] and more. One of their most promising applications has been optical fibre communications where they have enabled massively parallel ultrahigh capacity multiplexed data transmission [22,23]. Here, by using a new and powerful class of micro-comb called "soliton crystals" [11], we achieve unprecedented data transmission over standard optical fibre using a single integrated chip source. We demonstrate a line rate of 44.2 Terabits per second (Tb/s) using the telecommunications C-band at 1550nm with a spectral efficiency – a critically important performance metric - of 10.4 bits/s/Hz. Soliton crystals exhibit robust and stable generation and operation as well as a high intrinsic efficiency that, together with a low soliton micro-comb spacing of 48.9 GHz enable the use of a very high coherent data modulation format of 64 QAM (quadrature amplitude modulated). We demonstrate error free transmission over 75 km of standard optical fibre in the laboratory as well as in a field trial over an installed metropolitan optical fibre network. These experiments were greatly aided by the ability of the soliton crystals to operate without stabilization or feedback control. This work demonstrates the capability of optical soliton crystal micro-combs to perform in demanding and practical optical communications networks.**

The global optical fibre network currently carries hundreds of terabits per second per fibre every instant, with capacity growing at about 25% annually [24]. To dramatically increase bandwidth capacity, ultrahigh capacity transmission links employ massively parallel wavelength multiplexing (WDM) with coherent modulation formats [25,26], and in recent lab-based research, by using spatial division multiplexing (SDM) over multi-core or multi-mode fiber [27]. At the same time, there is a strong trend towards a greater number of shorter high-capacity links: whereas core long-haul (spanning 1000's km) communications dominated global networks ten years ago, now the emphasis has squarely shifted to metro-area networks (linking across 10's-100's km) and even data centres (< 10km). All of this is driving the need for increasingly compact, low-cost and energy efficient solutions, with photonic integrated circuits emerging as the most viable approach. The optical source is central to every link, and as such, perhaps has the greatest need for integration. The ability to supply all of the needed wavelengths with a single, compact integrated chip, replacing many parallel lasers, will offer the greatest benefits [28,29].

Micro-combs, optical frequency combs based on micro-cavity resonators, have shown significant promise in fulfilling this role, and have been successfully used as integrated chip sources for massively parallel ultrahigh capacity multiplexed data transmission [22,23,30]. This has been enabled by the ability to phase-lock, or mode-lock, the comb lines, a result of exploring novel oscillation states such as temporal soliton states including feedback-stabilized Kerr combs [23], dark solitons [30] and dissipative Kerr solitons (DKS) [22]. DKS states, in particular, have enabled transmission rates of 30 Tb/s for a single device and 55 Tb/s by combining two devices, using the full C and L telecommunication bands [22]. In particular, for practical systems, achieving a high spectral efficiency is critically important – it is a key parameter since it determines the fundamental limit of data-carrying capacity for a given optical communications bandwidth [25,26].

Here, we report record optical data transmission across standard fibre with a single integrated chip source. By using a novel and powerful class of micro-comb termed *soliton crystals* [11], realized in a CMOS (complementary metal-oxide semiconductor) compatible platform [2,3,31], we achieve a data line-rate of 44.2 Tb/s from a single source, along with a record spectral efficiency of 10.4 bits/s/Hz. We accomplish this through the use of a high modulation format of 64 QAM (Quadrature Amplitude Modulation), a low comb free spectral range (FSR) spacing of 48.9 GHz, and by using only the telecommunications C-band. We demonstrate transmission over 75 km of fibre in the laboratory as well as in a field trial over an installed network in the greater metropolitan area of Melbourne, Australia. Our results stem from the soliton crystal's robust and stable operation and generation and as well as its high intrinsic efficiency, all of which are enabled by an integrated CMOS-compatible platform.

Soliton crystals were so-named because of their crystal-like profile in the angular domain of tightly packed self-localized pulses within micro-ring resonators (MRRs) [11]. They are naturally formed in micro-cavities with appropriate mode crossings without the need for complex dynamic pumping and stabilization schemes that are often required to generate self-localised waves (described by the Lugiato-Lefever equation [32]). The key to their stability lies in their high intra-cavity power that is very close to that of spatiotemporal chaotic states [11, 33]. Hence, when emerging from chaotic states there is very little change in intracavity power and thus no thermal detuning or instability resulting from the 'soliton step' that makes resonant pumping more challenging [34]. It is this combination of intrinsic stability, ease of generation and overall efficiency that makes them highly attractive for demanding applications such as ultra-high capacity transmission beyond a terabit/s. These qualities have made soliton crystals highly successful as a basis for microwave and RF signal processing [ 35 - 50].

A schematic illustrating the soliton crystal optical structure is shown in Figure 1a), with the physical chip shown in Figure 1b) and the experimental setup for ultrahigh bandwidth optical transmission in Figure 1c) (also see Methods and Supplementary Material). The micro-resonator had an FSR spacing of 48.9 GHz and generated a soliton crystal with this spacing (approximately 0.4 nm) over a bandwidth of more than 80 nm when pumped with 1.8 W of continuous-wave (CW) power (in-fibre, incident) at 1550 nm. The soliton crystal micro-comb was generated by automatically tuning the pump laser to a pre-set value. The primary comb and generated soliton crystal states are shown in Figures 2 a&b. Figure 2c demonstrates the stability of the soliton crystal comb generation by showing a variation in individual tone powers of < +/- 0.9 dB over 10 different generation instances through wavelength sweeping (from 1550.300 to 1550.527 nm). This demonstrates the repeatability of turn-key micro-comb generation from this integrated CMOS-compatible device.

From the generated micro-comb, 80 lines were selected over the telecommunications C-band (32 nm wide, 3.95 THz window from 1536 – 1567 nm) which were then flattened with a spectral shaper (WaveShaper 4000S – see Methods). Next, the number of wavelengths were effectively doubled to 160 (equivalent to a comb spacing of 24.5 GHz) to optimize the spectral efficiency (spectrally useful content) and by using a single-sideband modulation scheme to generate odd/even decorrelated test channels (see Methods). We then combined a test band of six channels, with the remaining bands providing loading channels having the same odd-and-even channel structure. We used a record high order format of 64 QAM to modulate the entire comb at a symbol rate of 23 Giga-baud, resulting in the utilization of 94% of the available spectrum.

We conducted two transmission experiments, sending data over 75 km of single mode fibre in the laboratory as well as in a field trial across an installed metropolitan-area single-mode fibre network (see Supplementary Material) connecting the Melbourne City campus of RMIT and the Clayton campus of Monash University, spanning the greater metropolitan area of Melbourne. Spectra of the comb at key points are given in Figure 3 a-c. At the receiver, the signal was recovered using a common offline digital-signal processing (DSP) flow (see Methods). Figure 3d shows constellation diagrams for the signal at 194.34 THz. In back-to-back configuration (i.e. with transmitter directly connected to receiver) we measured signal quality ($Q^2$, from error vector magnitude) near 18.5 dB, dropping to near 17.5 dB when transmitting the fully modulated comb through the test links.

Figure 4a shows the transmission performance using the bit error ratio (BER) for each channel as a metric. Three scenarios were investigated: i) a direct connection between the transmitter stage to the receiver (back-to-back, B2B) and after transmission through ii) in-lab fibre and iii) over the field trial network. Transmission globally degraded the performance of all channels, as expected. As a performance benchmark, Figure 4a indicates a 20% soft-decision forward error correction (SD-FEC) threshold given at a BER of $4\times10^{-2}$ from a demonstrated code [51]. All results were below the given FEC limit, but since using SD-FEC thresholds based on BER is less accurate for higher order modulation formats and for high BERs [52], we additionally used generalized mutual information (GMI) to calculate the system performance. Figure 4b plots the GMI for each channel and its associated SE, with lines given to indicate projected overheads. We achieved a raw bitrate (line-rate) of 44.2 Tb/s, which translates to an achievable coded rate of 40.1 Tb/s (in B2B), dropping to 39.2 Tb/s and 39.0 Tb/s for the lab and field trial transmission experiments, respectively. These yielded spectral efficiencies of 10.4, 10.2 and 10.1 b/s/Hz (see Methods). This data rate represents an increase of nearly 50% (see Methods) for a single integrated device [22], while the spectral efficiency is 3.7 times higher. This is notable considering that the experiments were performed with the added penalty of full comb flattening (equalization, even though this is in fact not necessary [53]), and without closed-loop feedback or stabilization.

Our high transmission capacity and spectral efficiency are partly a result of the high conversion efficiency between the injected CW wave and the soliton crystal state [11, 33], which is reflected in the near absence of a soliton step (the jump in intra-cavity power that often occurs when generating coherent soliton states).

While our experiments were restricted to the C-band, the soliton crystal comb (Figure 2b) had a bandwidth exceeding 80 nm. The comb lines in the S and L bands (1500-1535 nm and 1565-1605 nm) could in principle be increased in power to enable transmission across all three bands, by varying the pump wavelength and power, tailoring dispersion and/or by other methods. Assuming similar comb quality, this would result in a three-fold increase in total data rate to 120 Tb/s from a single integrated device.

Micro-combs with lower FSRs would support even higher spectral efficiencies since the signal quality improves at lower symbol rates. However, this may come at the expense of a narrower overall comb bandwidth. In our demonstration, single-sideband modulation enabled the multiplexing of two channels onto a single light source, effectively halving the comb spacing while improving back-to-back performance as limited by transceiver noise. This approach is made feasible by the stable nature of the soliton crystal combs. On the other hand, electro-optic modulation can also be used to sub-divide the repetition rate of micro-combs, which would enable broad comb-bandwidths. While this would require locking the comb spacing to an external RF source, sub-MHz stabilization of comb spacings has been reported [54, 55]. Further, boosting the comb generation efficiency through laser cavity-soliton micro-combs [56] may provide a powerful path to improve signal quality and system capacity even further. Finally, for newly deployed links, our approach can readily be combined with space-division multiplexing using multi-core fibre [27, 57] which would result in data rates of many petabit/s from a single source.

In conclusion, we report record performance for ultra-high bandwidth optical transmission from a single integrated chip source by using soliton crystal micro-combs. This achievement is a result of the record low comb spacing combined with the stable, efficient, and broad bandwidth of the soliton crystal combs, all enabled by their CMOS compatible integration platform. Soliton crystal microcombs are intrinsically coherent, low noise, and can be initialised and maintained using standard open-loop control with off-the-shelf equipment. This work demonstrates their ability to support record ultrahigh bandwidth data transmission in practical and demanding environments.

# Methods

**CMOS compatible micro-comb source**

*Micro-ring resonator*

The micro-ring resonator (MRR) for comb generation was fabricated using CMOS compatible processes [4,30,31] with doped silica glass waveguides, which features low linear loss (~0.06 dB.cm$^{-1}$), a moderate nonlinear parameter (~233 W$^{-1}$.km$^{-1}$), and negligible nonlinear loss that leads to an ultra-high nonlinear figure of merit. The MRR has a cross-section of 3×2 μm and a radius of ~592 μm, yielding an FSR of 48.9 GHz and a Q factor >1.5 million. The dispersion of the TM mode was designed to be anomalous in the C band with a jump at ~ 1552 nm brought about by the mode crossing. The bus waveguides of the MRR were directed to on-chip mode converters, then coupled to a single-mode fibre array, with a fibre-chip coupling loss of ~0.5 dB per facet.

While statistical studies of fabrication yield are outside the scope of this work, we note that our platform is fully CMOS compatible using stepper mask aligners on full wafers [58]. Further, our low index contrast (core index = 1.7), results in larger waveguide dimensions which in turn makes them less sensitive to fabrication error. Our typical yields for FSR and Q factor are extremely high – well above 90%, and mode-crossings do not pose a particular challenge. We have fabricated many soliton crystal devices [33] with high reproducibility. The discovery that mode crossings provide a new and powerful approach towards achieving robust or even deterministic generation of microcombs shows that further engineering of these structures remains an important and highly beneficial challenge that will yield new functionality.

*Soliton crystal micro-comb generation*

The micro-comb used in the study was generated from the doped silica double-bus micro-ring resonator described above, packaged with a fibre array connection to all four device ports. We pumped the ring with a CW external cavity laser (Yenista Tunics – 100S-HP) at an output power of 15 mW, which was then amplified to 1.8W in a polarization maintaining erbium doped fibre amplifier (EDFA) (Pritel PMFA-37). Only the TM mode of the resonator oscillated in a soliton crystal state, hence the pump polarization was tuned to match this mode. As indicated in Figure 1a, we inserted the pump light into the 'through' port and collected light from the corresponding 'drop' port. The MRR chip was mounted on a Peltier cooler, monitored by a standard NTC temperature sensor. The temperature was maintained with a thermo-electric cooler (TCM-M207) at 25$^o$C, within 0.1$^o$C of accuracy. The laser was set to standard running mode, with no extra steps made to stabilise the output frequency. Soliton crystal generation was achieved by automated wavelength tuning (described further in the Supplementary Material), in turn reducing the system complexity compared to other micro-comb generation schemes (see [6] and references within). We measured the internal conversion efficiency of our soliton crystals to be 42% for the whole spectrum, and 38% when selecting the 80 lines over the C-band, highlighting that over 90% of our available comb power is compatible with standard C-band equipment (see Supplementary Material S.M.2 for detail).

The generated soliton crystal micro-comb was then flattened in two stages by two independent programmable optical filters (Finisar WaveShaper 4000S). The WaveShapers had an insertion loss of 5dB each, in addition to any variable attenuation. The first had a static filter shape set to equalize each comb line to within about 1 dB of each other, to coarsely match the generic shape of the soliton crystal state we chose to use. The second programmable filter was set each time that a new soliton crystal state was initiated, to equalize the comb line powers to within < 1 dB of each other, although we note that it was often unnecessary to change the filter profile when generating a new soliton crystal. Spectral shaping in a WDM transceiver using a comb source involved minimal extra complexity since only the addition of attenuators after the WDM demultiplexer were required to route each comb line to a separate modulator. The comb was then amplified by a further polarization maintaining EDFA (Pritel PMFA-20-IO), before being divided for modulation. Prior to modulation, the optical signal-to-noise ratio (OSNR) of the individual comb lines was > 28 dB (see Supplementary Material).

The nonuniform spectrum of soliton crystal combs has been considered as a drawback, and so for this reason, as well as to facilitate easier comparison with prior work using micro-combs, we ensured that the optical frequency comb was flattened such that all lines were of equal power. Comb flattening in fact is not necessary either in our experiments or other micro-comb demonstrations (e.g. [22,30]), since all comb lines are typically wavelength demultiplexed into separate waveguides and sent to separate modulators. It is then straightforward to adjust the comb line power by variable attenuators, amplifiers or even by varying the RF drive amplitude to the modulators. In fact, we expect better performance without comb flattening since the higher power comb lines would need less attenuation and/or amplification before modulation, resulting in a higher OSNR, while the lower power comb lines would have essentially the same performance as reported here. Furthermore, using the raw spectrum would avoid the loss of the extra Waveshaper. Therefore, avoiding flattening (working with the raw spectrum) would in fact yield even higher system performance.

**Systems experiment**

*Transmitter stage*

The detailed experimental setup is shown in the supplementary material Figure SM1. The transmitter used 3 separate complex Mach-Zehnder modulators to provide both odd and even test bands, as well as a loading band. The comb lines for each of these bands were split using another programmable filter (Finisar WaveShaper 4000S) and were then amplified before modulation. Three tones separated by 98 GHz around the selected test frequency were directed to two separate modulators (Sumitomo Osaka Electric Company New Business T.SBXH1.5-20 ). The two modulators were driven at a symbol rate of 23 Gbd, providing a per sub-band line rate (i.e. excluding overheads) of 23 Giga-symbols/s x 6 bits/symbol x 2 polarizations = 276 Gb/s. The sub-bands were shifted by 12 GHz from the optical carrier, with one modulator providing a sideband down-shifted from the optical carrier, and the other an up-shifted sideband. This enabled higher fidelity modulation than simple 46 Gbd single-carrier modulation, given the transceiver noise limitations we had in our system. The odd and even bands were decorrelated by adding a delay with an extra length of optical fibre in the 'even' path. A third modulator (Covega Mach-40 806) was used to modulate the loading bands which consisted of a two Nyquist sub-carrier modulation scheme to mimic the structure of the odd and even test bands. The two bands were driven by pairs of the positive and negative differential outputs of the AWG (Keysight M8195A, 65 GSa/s, 25 GHz bandwidth), while the loading channels were driven by a separate independent output pair. The modulating waveforms were set to provide 64 QAM signals, pulse shaped by a 2.5% roll-off RRC filter, running at 23 Gigabaud. On a 49 GHz grid, this provided a 94% spectral occupancy. The modulator optical outputs were each passed through a polarization maintaining 3 dB power splitter, one output being delayed by a few meters of optical fibre and then rotated by 90º using a polarization beam splitter/combiner. This provided emulation of polarization multiplexing by delay de-correlation. The odd, even and loading bands were all de-correlated from each other by means of different fibre delays of a few meters. The odd and even channels were passively combined with a 3-dB power splitter, to maintain the pulse shape of the central channels. The combined test and loading bands were multiplexed by a further programmable filter (Finisar WaveShaper 4000S). The roll-off of the filters from this device did affect the outer channels of the test band and the neighbouring channels in the loading channels. After multiplexing the fully modulated comb was amplified to a set launch power. The Tx DSP is described in the supplementary material.

*Field-trial link*

The physical fibreoptic network geography is shown in Fig. SM2 and the schematic layout in Fig. SM1 (see Supplementary Material). The transmission link was comprised of two fibre cables connecting labs at RMIT University (Swanston St., Melbourne CBD) and Monash University (Wellington Rd, Clayton). These cables were routed from the labs access panels, to an interconnection point with the AARNet's fibre network. The fibre links are a mix of OS1 and OS2 standard cables and include both subterranean and aerial paths. There is no active equipment on these lines, providing a direct dark fibre connection between the two labs. The total loss for these cables was measured to be 13.5 dB for the RMIT-Monash link and 14.8 dB for the Monash-RMIT paths. The cable lengths as measured by OTDR were both 38.3 km (totalling 76.6 km in loop-back configuration). At

Monash, an EDFA was remotely monitored and controlled using a 1310 nm fibre-ethernet connection running alongside the C-band test channels. The comb was amplified to 19 dBm before launch, at Monash, and upon return to RMIT.

*Receiver stage*

The receiver stage architecture is shown in Figure SM1. Before photo-detection, the signal was filtered by a programmable optical filter (Finisar WaveShaper 4000S) set to a 35 GHz passband, in order to select the channel to be measured. The 35 GHz passband was found to be an optimal setting in experiment (see Supplementary Materials for more detail). The output of the filter was amplified to approximately 10 dBm before being directed into a dual-polarization coherent receiver (Finisar CPDV1200, 43 GHz bandwidth). A local oscillator was provided by an Agilent N7714A laser tuned close to the comb line of interest, at 16 dBm of output power. The photo-detected signals were digitized by the 80-giga-samples-per second (GSa/s), 33-GHz bandwidth inputs of a Keysight oscilloscope (DSO-Z504A, 50 GHz, 160 GSa/s). The digitized waveforms were forwarded to a PC for offline digital signal processing. The digital signal processing flow started with renormalization, followed by overlap-add chromatic dispersion compensation, then a spectral peak search for frequency offset compensation, followed by frame synchronization using a short BPSK header, before final equalization. As the specific fibre types used along the link are not well known, the level of chromatic dispersion compensation was estimated through analysing the header correlation peak height. Equalization occurred in two stages, with a training-aided least-means-squared (LMS) equalizer performing pre-convergence, the taps of which were sent to a blind multi-modulus equalizer. After equalization, a maximum-likelihood phase estimator was used to mitigate phase noise, before the signal was analysed in terms of BER, EVM and GMI. Further details are included in supplementary material.

*Notes on calculation of system performance metrics*

After signal reconstruction using DSP, we measured the system performance based on three separate metrics: bit-error rate (BER), error-vector magnitude (EVM) and generalized mutual information (GMI).

BER is measured by decoding a 1.1-Mbit-long random bit sequence that was grey-coded onto the 64-QAM constellation. As such, a BER of $9 \times 10^{-5}$ provides 100 errors.

Error-vector magnitude provides an alternative metric, which is directly related to the effective signal-to-noise ratio (SNR) measured at the receiver in the presence of uniform Gaussian noise. We use EVM to calculate signal quality factor ($Q^2$ [dB]) in Figure 3 as $20\log_{10}(1/EVM^2)$, with $EVM = 1/SNR^{0.5}$.

In systems employing higher-order modulation formats, generalized mutual information (GMI) provides a more accurate measure of system performance than taking BER and assuming a certain soft-decision forward error correction threshold [52]. We use GMI to provide the key performance figures in this demonstration (i.e. net data rate and spectral efficiency). In this case, achievable capacity (b/s) is calculated as the sum of individual channel GMIs (b/symbol) and multiplying by the symbol rate (symbols/s).

Spectral efficiency (b/s/Hz) can also be calculated from GMI, by taking the mean of the channel GMIs (b/symbol), multiplying by the symbol rate (symbols/second) and dividing by the channel spacing (Hz).

*Performance Comparison Summary*

Table 1 summarizes key results from the literature comparing the various system performance metrics for demonstrations based on a single integrated source and over standard fibre (or calculated on a per-mode basis for multicore fibre). Previous to this work, the best result (per core) was from [22]. In that work, a single microcomb was able to support 30.1 Tb/s over the C and L bands, when using a standard tuneable laser coherent receiver. This is the benchmark result that we compare our results to in the main text, since it is not only the best published result using a single micro-comb, but it closely resembles our experiment (single micro-comb at transmitter, single tuneable laser at the receiver as a local oscillator). Note that our system uses less than half the spectrum of [22], while substantially exceeding its data rate, due to our much higher spectral efficiency (3.7x higher). High modulation formats have also been achieved with dark solitons [30], yet at a lower overall data rate, primarily due to the high comb line spacing that significantly limits the spectral efficiency. The work of [27] used a comb generator based on a benchtop pulsed seed laser source combined with waveguide spectral broadening. To provide a fully integrated system, this source would need to be on-chip. The focus in that experiment was using novel, proprietary multi-core fibre to achieve a 30-fold increase in bandwidth over the fibre in this spatially multiplexed system, to reach 0.66 Petabits/s. On a per-mode basis, ref. [27] yields 25.6 Tb/s/mode, a lower per-mode capacity than this work and that of [22]. We note that both our approach and that of [22] are able to take advantage of SDM techniques to scale the overall bandwidth by using multi-core fibre. We provide further comparisons in the supplementary material.

**Table 1. Key systems performance metrics.**

| Line Rate | Net Rate | Format | Spectral Efficiency | Transmission | Source |
|---|---|---|---|---|---|
| 30.1 Tb/s | 28.0 Tb/s | 16QAM | 2.8 b/s/Hz | 75 km SMF in-lab | Ref. [22] |
| 4.8 Tb/s* | 4.4 Tb/s | 64QAM | 1.1 b/s/Hz* | 80 km SMF in-lab | Ref. [30] |
| 25.6 Tb/s* | 22.0 Tb/s* | 16QAM | 3.2 b/s/Hz* | 9.6 km, 30-core fibre | Ref. [27] |
| 44.2 Tb/s | 40.1 Tb/s | 64QAM | 10.4 b/s/Hz | B2B (0 km) | This work |
| 44.2 Tb/s | 39.2 Tb/s | 64QAM | 10.2 b/s/Hz | 75 km SMF in-lab | This work |
| 44.2 Tb/s | 39.0 Tb/s | 64QAM | 10.1 b/s/Hz | 76.6 km SMF installed | This work |

The yellow highlighted results from [27] used chip-based of a non-integrated benchtop pulsed laser source, and are shown for reference. For this space division multiplexing (SDM) demonstration we quote numbers that are per spatial mode.

'*' indicates that this figure was not directly provided in the reference, and so is inferred from data provided.

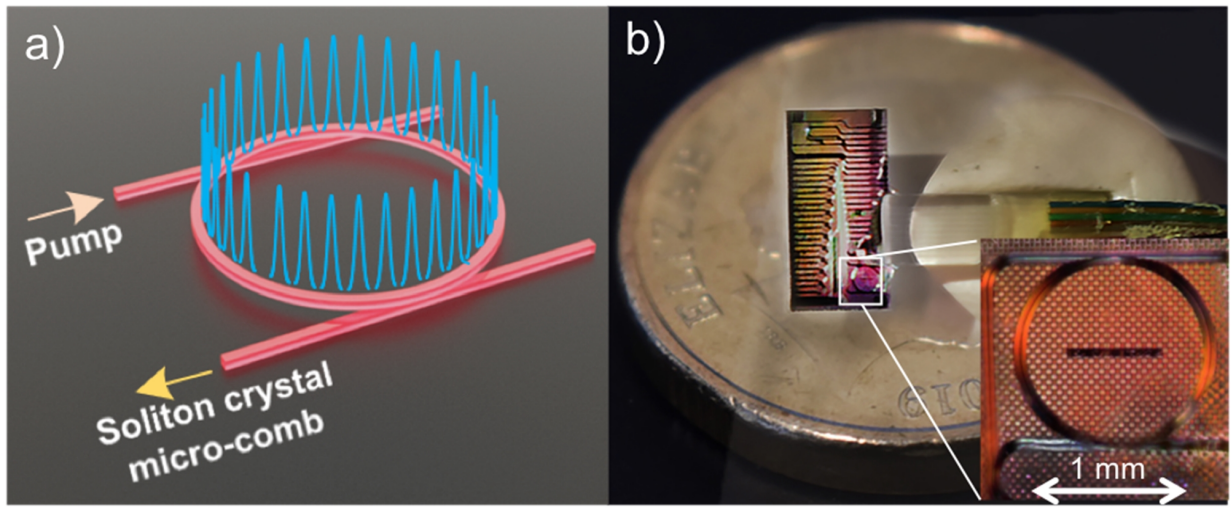

**Figure 1. Conceptual diagram of a soliton crystal micro-comb communications experiment.** a) Illustration of the soliton crystal state used in this paper. We infer from the generated spectrum that the state was a single temporal defect crystal across the ring. The state had a characteristic 'scalloped' micro-comb spectrum, corresponding to the single temporal defect crystal state illustrated over the ring. b) Photograph of the fibre-optic packaged micro-ring resonator chip used for soliton crystal generation. The full chip is 5 mm x 9 mm, of which we use devices and access waveguides on approximately ¼ of the area. The AUD $2 coin (20.5 mm diameter) shown for scale is similar in size to a USD nickel or a 10 Euro cent coin. Inset is a microscope image of the ring resonator element, with a scale bar. Visible distortions are due to an overlayer of glue from the fibre array. c) Experimental set-up. A CW laser, amplified to 1.8W, pumped a 48.9 GHz FSR micro-ring resonator, producing a micro-comb from a soliton crystal oscillation state. The comb was flattened and optically demultiplexed to allow for modulation, and the resulting data optically multiplexed before the subsequent transmission through fibres with EDFA amplification. At the receiver, each channel was optically demultiplexed before reception. ECL, edge-coupled laser, WSS wavelength-selective switch. Rx receiver.

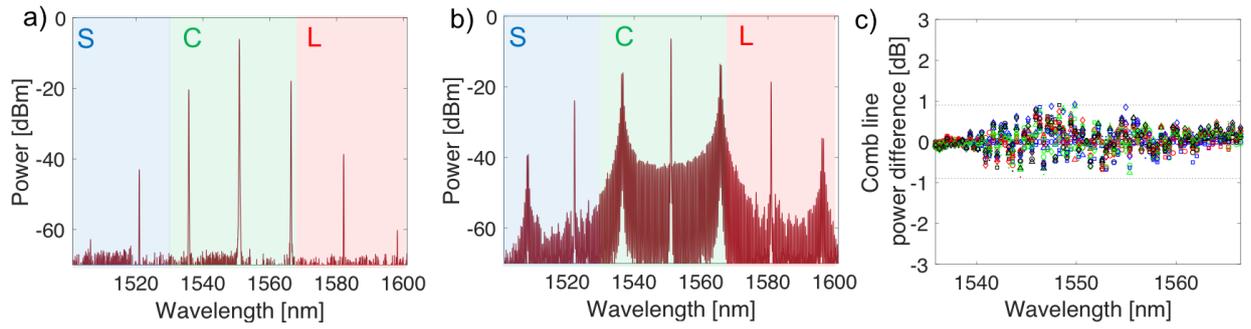

**Figure 2. Soliton crystal generation.** To generate a soliton crystal, a laser is slowly tuned from the red side of a resonance to a pre-set wavelength. a) A primary comb (Turing pattern) is initially generated as the laser is tuned into resonance with the ring. b) Spectrum of the soliton crystal oscillation state used for experiments. The state had a characteristic 'scalloped' micro-comb spectrum, corresponding to the single temporal defect crystal state illustrated over the ring. At the pre-set wavelength, a soliton crystal forms, with spectral features based around the primary comb lines. The state that we use provides comb lines over most of the optical communications C-band. c) Soliton crystal comb line power difference for 10 independent crystal generation instances (different symbols indicate distinct generation instances). Comb line powers remain within +/- 0.9 dB of the initial spectrum, indicating reliable generation of the desired soliton crystal state.

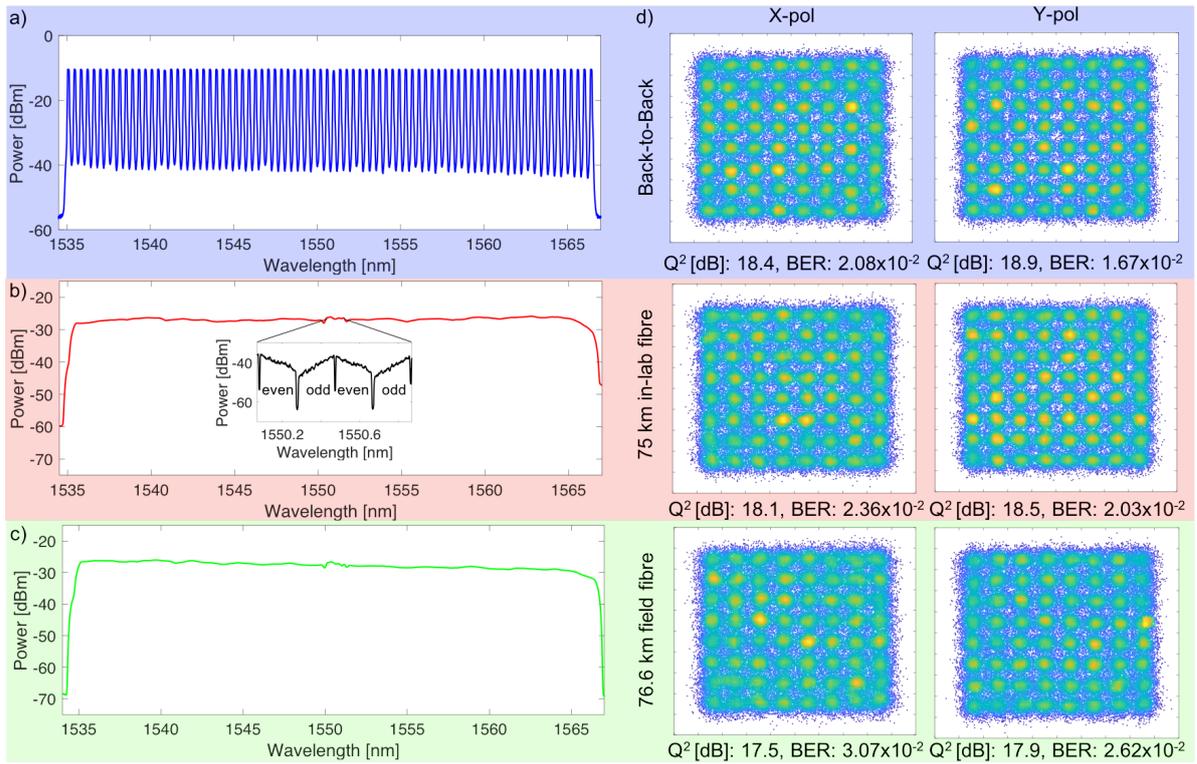

**Figure 3.** a-c) Spectra of the soliton crystal frequency comb after flattening (a), modulation and transmission through either 75 km spooled in-lab fibre (b) or through the field-trial link (c). The spectrum (a) is measured with 12.5 GHz resolution to resolve the individual comb lines, while (b) and (c) are plotted at 50 GHz resolution to illustrate average channel powers. Flattening equalised the comb line power to within 1 dB. After modulation and amplification, the channels were shaped by the EDFA gain spectrum. The inset in (b) depicts the test channel spectra captured with a 150 MHz resolution optical spectrum analyser (Finisar WaveAnalyzer), highlighting the odd and even sub-bands modulated onto each comb line in the test band. d) Constellation diagrams for a comb line at 193.4 THz (1550.1 nm) for both X- and Y-polarization channels. 'Back to back' denotes the transmitter directly connected to the receiver, '75 km in-lab fibre' indicates reception after transmission through 75 km of spooled fibre inside the lab (as per b), while '76.6 km field fibre' denotes reception after transmission through the field-trial link (as per c). BER and $Q^2$ related to the constellations are noted on each.

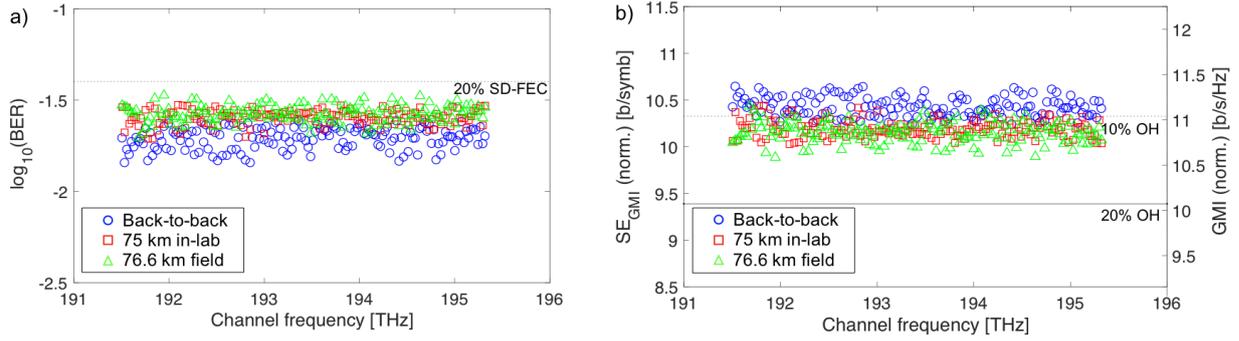

**Figure 4**. a) BER for each comb line. Blue circles points indicate performance of channels in B2B configuration, red squares dots are for performance after transmission through 75 km of in-lab spooled fibre, while green triangles are after transmission through the 76.6 km installed metropolitan-area fibre link. An indicative FEC threshold is given at $4\times10^{-2}$, corresponding to a pre-FEC error rate for a 20% soft-decision FEC based on spatially-coupled LDPC codes [25] (dashed line). After transmission, all channels were considered to be error-free, b) GMI and spectral efficiency measured for each comb line. GMI was calculated after normalization to scale measured constellations in order to account for received signal-to-noise ratio (SNR). Lines are for 20% and 10% overheads. Spectral efficiency was derived from GMI, and the ratio of symbol rate to comb spacing. GMI indicates a higher overall capacity than BER with the indicated SD-FEC threshold, as GMI assumes the adoption of an ideal code for the system. For B2B, GMI (SE) varied between 11.3 b/symb. (10.6 b/s/Hz) and 10.9 b/symb. (10.3 b/s/Hz). After in-lab fibre transmission, the achievable per-channel GMI (SE) varied between 11.0 b/symb. (10.4 b/s/Hz) and 10.7 b/symb. (10.1 b/s/Hz), with the same range observed for the installed field-trial fibres. We estimate the overall capacity from the sum of the GMIs, multiplied by the symbol rate.

**Acknowledgments**

We gratefully acknowledge support from Australia's Academic Research Network (AARNet – aarnet.edu.au) for supporting our access to the field trial cabling through the project Australian Lightwave Infrastructure Research Testbed (ALIRT), and in particular Tim Rayner, John Nicholls, Anna Van, Jodie O'Donohoe and Stuart Robinson.

A.M. and B.C. acknowledge funding through the Australian Research Council's Discovery and Linkage Infrastructure schemes (DP190102773, LE170100160). B.C. thanks Dr. Tobias Eriksson (NICT, Japan) for helpful advice regarding GMI calculations.

R.M. acknowledges funding by the Natural Sciences and Engineering Research Council of Canada (NSERC) through the Strategic, Discovery, and Acceleration Grants Schemes, by the MESI PSR-SIIRI Initiative in Quebec, by the Canada Research Chair Program.